# Nonreciprocal transmission in a photonic-crystal Fano structure enabled by symmetry breaking

Yi Yu[†], Yaohui Chen, Hao Hu, Weiqi Xue, Kresten Yvind and Jesper Mork

*DTU Fotonik, Department of Photonics Engineering, Technical University of Denmark, DK-2800 Kongens Lyngby, Denmark*

[†]E-mail address: yiyu@fotonik.dtu.dk

**Abstract**

Nanostructures that feature nonreciprocal light transmission are highly desirable building blocks for realizing photonic integrated circuits. Here, a simple and ultra-compact photonic-crystal structure, where a waveguide is coupled to a single nanocavity, is proposed and experimentally demonstrated, showing very efficient optical diode functionality. The key novelty of the structure is the use of a Fano resonance in combination with spatial symmetry breaking and cavity-enhanced material nonlinearities to realize non-reciprocal propagation effects at ultra-low power and with a good wavelength tunability. The nonlinearity of the device relies on ultrafast carrier dynamics, rather than the thermal effects usually considered, allowing the demonstration of non-reciprocal operation at a bit-rate of 10 Gbit/s with a low energy consumption of 4.5 fJ/bit.

**Introduction**

Nanostructures that break time-reversal symmetry are highly desirable building blocks for next generation all-optical signal processing circuits, providing analogues to the electrical diodes and transistors that have enabled today's computer and information technology. Diverse configurations have been proposed in order to realize asymmetrical light propagation, including devices based on magneto-optic effects [1-4], indirect interband photonic transitions [5-9], opto-acoustic effects [10], nonlinear gain or absorption [11, 12], parametric processes [13] as well as thermal nonlinearities [14-16]. Although these approaches have their own advantages, on-chip integration is still difficult to achieve since the structures demonstrated so far suffer from high complexity, costly fabrication, high energy consumption, small bandwidth or large footprint.

In Refs [14-16], passive optical diodes based on optical nonlinearity in cascaded high quality-factor cavities were proposed and demonstrated, showing a large nonreciprocal transmission ratio (NTR) of 40 dB [16]. These devices do not require any externally applied fields, but in order to achieve large NTR, the resonant frequencies of the cavities need to be accurately matched [14, 16], requiring elaborate tuning approaches to compensate for the resonance mismatch introduced upon fabrication. In addition, the energy consumption is relatively large and the devices are narrow-band, i.e. a large NTR can be only achieved within a bandwidth of 0.1 nm or less. Furthermore, these diodes are actually reciprocal when light is applied from both directions simultaneously, making them unsuited as isolators for continuous-wave (CW) lasers. Non-reciprocal structures realized by optical nonlinearities, may, however, be used to isolate pulsed sources, like mode-locked lasers, provided the dynamics of the diode is fast enough, a requirement not met when basing the non-reciprocity on thermo-optic effects as in Refs [14-16].

Here, we demonstrate a simple and ultra-compact photonic crystal (PhC) cavity-waveguide structure that exploits spatial symmetry breaking in combination with a Fano resonance and ultrafast carrier dynamics to realize non-reciprocity at low-power levels and with a good wavelength tunability. In contrast to previous work utilizing multi-cavity schemes [14-16], our structure relies on the coupling of a single, discrete-resonance, H0 nanocavity to the (longitudinal) mode continuum of a line-defect waveguide, alleviating the need to accurately control the nanocavity resonance. By introducing a partially transmitting element in the waveguide, the transmission spectrum of the structure exhibits an asymmetric Fano spectrum, offering a sharp transmission change within a narrow wavelength range. This Fano resonance in combination with efficient cavity-enhanced nonlinearity enables the structure to achieve simultaneously large NTR (35 dB), ultra-low power operation (-6.5 dBm) and tunability (1 nm). Furthermore, while previous work [14-16] was limited to investigations of the static behavior, we here demonstrate the successful operation of the diode under dynamical operation. For realistic input pulse sequences at a bit-rate of 10 Gbit/s, the diode is thus found to be non-reciprocal at energies as low as 4.5 fJ/bit.

**Device structure**

Fig. 1(a) shows a scanning electron microscope image of one of the fabricated devices based on an InP PhC air-embedded membrane structure, which consists of a line defect waveguide coupled to a point defect cavity (H0-type) featuring a very small mode volume. The device is fabricated using a combination of electron-beam lithography, reactive-ion etching and selective wet-etching, with the detailed fabrication processes described in Ref. [17]. In contrast to ordinary side-coupled systems [18], a partially transmitting element (PTE) is introduced in the waveguide above the nanocavity [19] by etching a blockade-hole in the waveguide. The transmission coefficient $t_B^2$ of this partial blockade may be controlled via the size (radius $R_B$) of the hole. We henceforth denote light transmitted from port 1 (port 2) to port 2 (port 1) as forward (backward) propagating. The mirror symmetry of the structure is broken by displacing the PTE by one lattice constant towards port 1. The device is equipped with mode adapters to facilitate out coupling [20].

Asymmetry between forward and backward transmission originates from two key elements. First, due to the broken mirror symmetry, the coupling rate, $\gamma_1$, between port 1 and cavity is different from the rate, $\gamma_2$, between port 2 and cavity. This means that the energy excited in the cavity depends on whether one considers forward or backward transmission, causing different nonlinear shifts of the cavity resonance for the two directions and leading to nonreciprocal light transmission. Notice that the difference between the two decay rates is due to interference effects, i.e., the structure does not rely on an absorptive element, distinguishing it from devices exploiting saturable absorption [12]. The second element is that the PTE allows controlling the amplitude of the continuum-path. This makes part of the injected light propagate through the localized cavity resonance, while another part of the light propagates directly through the waveguide, causing interference at the output. Thus, instead of displaying an ordinary symmetric Lorentzian spectrum, the transmission of such structure exhibits a Fano line [19, 21] with an asymmetric spectrum as illustrated in Fig. 1(b,i). Compared to its Lorentzian counterpart (Fig. 1(b,ii)), this Fano line features a much larger transmission change within a narrower wavelength range, thus enabling large NTR with a small input power variation. For the fabricated structure, a large on-off contrast of 40 dB was achieved, cf. Fig. 1(b,i).

Fig. 1(b) also shows theoretical results obtained using temporal coupled mode theory [22, 23]. The linear transmission of the investigated waveguide-cavity system (Fig. 1(c)) can be expressed as (see supplementary information, section A):

$$t(\omega) = \frac{t_B(\omega_0 - \omega) \pm \sqrt{4\gamma_1\gamma_2 - t_B^2(\gamma_1 + \gamma_2)^2} - jt_B\gamma_v}{j(\omega_0 - \omega) + \gamma_t}. \quad (1)$$

where $t_B$ and $r_B = \sqrt{1-t_B^2}$ are the amplitude transmission and reflection coefficients of the PTE. The intrinsic and total loss rates, $\gamma_v$ and $\gamma_t = \gamma_1 + \gamma_2 + \gamma_v$, can be related to the cavity intrinsic and total quality-factors, $Q_v$ and $Q_t$, as $\gamma_v = \omega_o / 2Q_v$ and $\gamma_t = \omega_o / 2Q_t$. By fitting the measured transmission spectrum to the expression Eq. (1) using least squares method, the structure is found to have a resonant wavelength at 1575.8 nm with a $Q_v$ of $1.2 \times 10^5$ and $Q_t$ of $1.35 \times 10^3$.

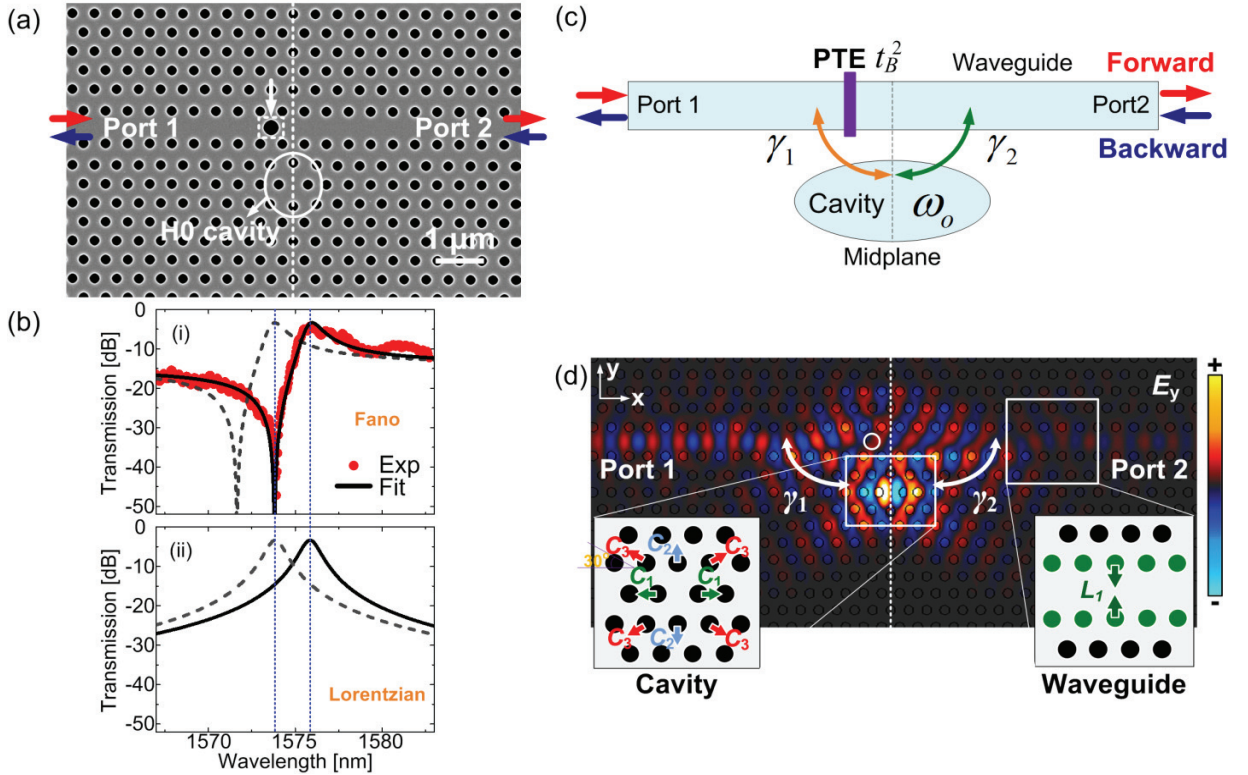

**Figure 1 Cavity-waveguide system with broken mirror symmetry.** (**a**) Scanning electron microscope image of the fabricated sample based on an InP PhC air membrane structure. A PTE (dashed square) is formed by etching one blockade-hole in the waveguide. The structure is made asymmetric with respect to the mid-plane (vertical dashed line) by displacing the PTE one lattice constant towards port 1. (**b**) (i) Measured linear transmission (Fano) spectrum (red dots), and its corresponding theoretical fit (black solid line). The spectrum is normalized so that its off-resonance transmission equals $t_B^2$. (ii) Corresponding theoretical Lorenztian spectrum with the same total and intrinsic quality-factors as those of the Fano spectrum. Much larger transmission contrast can be achieved for the Fano resonance compared to its Lorentzian counterpart for the same resonance shift (indicated by the vertical blue dashed lines). (**c**) Schematic of the investigated system. A signal at frequency $\omega$ passes through a waveguide that is side-coupled to a cavity with resonance frequency $\omega_0$. (**d**) PhC cavity-waveguide structure overlapped with $E_y$ field component of the cavity mode obtained by FDTD simulations. The insets illustrate the detailed PhC structure: an H0 cavity is formed by displacing 8 air holes by the amounts $C_i$ ($i = 1\sim3$). The waveguide is of the standard W1 type (defined as the removal of one row of air holes) except for the innermost row of air holes (green air holes) being shifted by a distance $L_1$ towards the waveguide center. The air hole shifting directions are indicated by the arrows. The white circle indicates the blockade hole with a radius of $R_B$.

The detailed structure design is shown in Fig. 1(d). Several air hole positions are tuned around the cavity in order to optimize the cavity intrinsic quality-factor. The waveguide dispersion is tailored by shifting the innermost row of air holes towards the waveguide center to make the frequency of the waveguide fundamental mode overlap with the cavity mode. The mirror symmetry of the structure is broken by displacing the PTE by one lattice constant towards port 1. Fig. 1(d) also shows the $E_y$ field component distribution, calculated using three-dimensional finite difference time domain (FDTD) simulations upon exciting the cavity. The figure illustrates that different amounts of light are coupled out via port 1 and 2, representing the difference between the coupling rates $\gamma_1$ and $\gamma_2$. The specific parameters for the design are: air hole radius $R = 0.253a$, $R_B = 1.5R$, membrane thickness $h = 0.76a$, $C_1 = 0.16a$, $C_2 = 0.15a$, $C_3 = 0.074a$, $L_1 = 0.12a$ where $a$ is the lattice constant. From the FDTD simulations, the PhC structure exhibits a small linear mode volume of only $0.25(\lambda/n)^3$ with $t_B = 0.19$ and $\gamma_1/\gamma_2 = 6.5$, which is estimated by comparing the electromagnetic flux leaking into ports 1 and 2 when exciting the corresponding eigenmode of the cavity.

**Characterization of static properties**

The transmission spectrum of the fabricated device is measured using a tunable CW laser source in combination with an optical spectrum analyzer. The input light is TE-polarized (electric field in the plane of the PhC slab) and is coupled in and out of the device using lensed single mode fibers. An infrared camera (Xeva-1.7-320) mounted on a microscope is used to image the far-field scattered light from the top of the sample [17]. To minimize heating and temperature fluctuations, the chip is coated with a thin layer of water and is mounted on a copper bench with its temperature controlled by a thermoelectric controller. The device thermal properties may be further improved by encapsulating the device in silica which has a higher thermal conductivity [24].

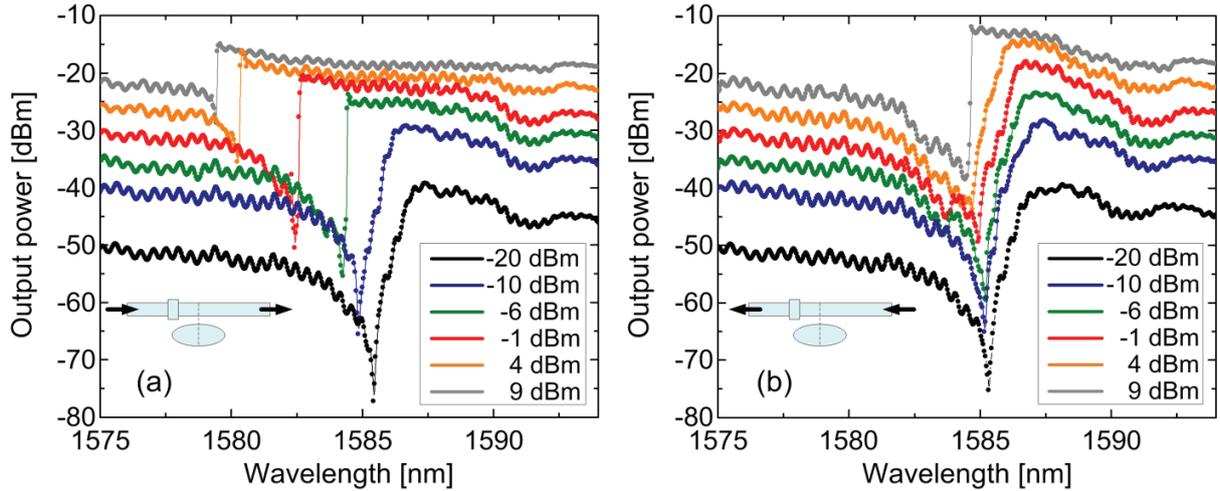

**Figure 2 Measured transmission spectra.** (**a**) Forward and (**b**) backward transmission spectra for different values of the input power. The measured output power includes both the coupling losses (between the tapered fibers and the PhC mode adaptors) and the propagation loss in the device. The insets show the light transmission directions.

Fig. 2 shows the measured transmission spectra. Since the chip is coated with water, the original cavity resonance shifts to longer wavelength (1586.6 nm) and the on-off contrast is reduced, reflecting a reduction of the intrinsic quality-factor. Nevertheless, a large on-off contrast of 35 dB is achieved. For sufficiently low input power (-20 dBm), the measured forward and backward transmission spectra are nearly identical and the measurements agree very well with linear theory, where the resonance frequency is not perturbed by the energy build-up in the cavity. The oscillations in the measured spectra with a period of 0.4-0.5 nm are Fabry-Perot fringes due to reflections at the waveguide end-facet and the cavity. As the input power increases, the spectra change significantly, reflecting a nonlinear change of the resonance frequency. The observed effects are consistent with a reduction of the refractive index due to a combination of plasma effects and band-filling caused by free carriers generated via two-photon absorption [25-29]. The abrupt changes of output power with wavelength occur due to the existence of a spectral region of bistability located on the blue side of the resonance. Since the tunable laser was scanned from shorter to longer wavelengths in Fig. 2, the transmission normally follows the lower branch of the bistability curve. Upon increasing the input power, both forward and backward transmission spectra broaden towards shorter wavelengths. However, this effect is much larger for the forward transmission, which is attributed to the larger cavity excitation and thus nonlinearity caused by the larger coupling rate between port 1 and cavity. These interpretations are supported by theoretical modeling, showing very good agreement between experiment and theory (see supplementary information, section B1). At higher input power levels, the transmission is observed to saturate, which is ascribed to the blue-shifting of the cavity induced by the pump light, which implies that a smaller fraction of the pump light is coupled into the cavity. In addition, nonlinear losses due two two-photon and free-carrier absorption increase with pump power.

Next, the transmission properties are characterized in dependence of input power and signal detuning, $\delta_\lambda = \lambda_0 - \lambda_s$, where $\lambda_0$ and $\lambda_s$ are the cavity resonance wavelength and the input signal wavelength, respectively.

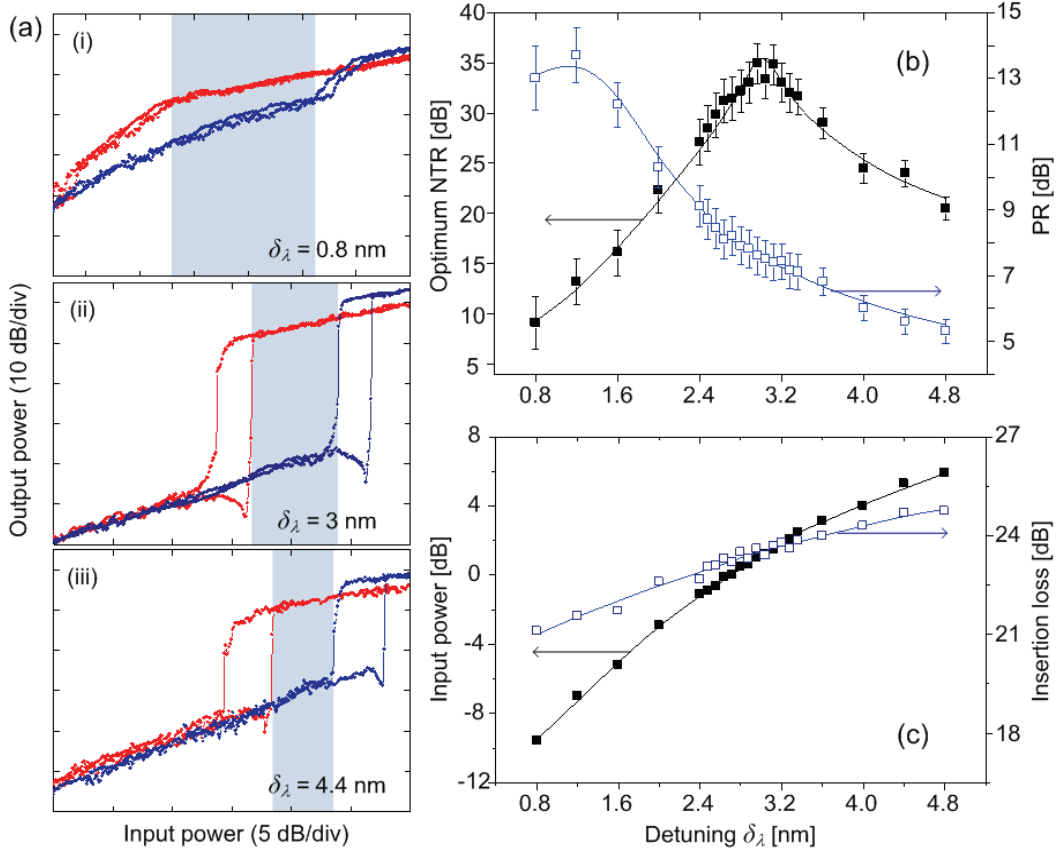

**Figure 3 Nonreciprocal transmission under CW illumination.** (**a**) Examples of measured output power versus input power in forward (red) and backward (blue) direction for signal detunings of (i) 0.8 nm, (ii) 3 nm and (iii) 4.4 nm. The shaded regions represent the stable regions where the NTR and power range can be extracted. (**b**) The optimum NTR (black squares) and power range (blue squares) as a function of signal detuning. (**c**) Light power at input fiber (black squares) and insertion loss (blue squares, including coupling and propagation losses) in the forward direction, corresponding to the optimum NTR depicted in (**b**), as a function of signal detuning. The lines are cubic spline fits. The error bar reflects the measurement uncertainties, with the average corresponding to 5 consecutive measurements.

Fig. 3(a) shows the forward and backward transmission measured by sweeping the input power first from low to high values and subsequently from high to low values. Bistability occurs for relatively large, positive, values of the signal detuning. The local dips observed in the lower branches of the bistability curve are a direct evidence of the Fano line shape [30]. As the signal blue shifts far away from the resonance wavelength, the unstable region of the bistable curve (where the output power may fluctuate between the values corresponding to the lower and upper branch) broadens, and the power difference between the upper and lower branches first increases and then decreases. For detunings larger than 5 nm, the bistability curve is difficult to measure due to the limited power available. It is clear from the measurements that bistability is achieved at much lower input power levels in the forward direction than in the backward direction. The transmission thus becomes nonreciprocal above a threshold incident light power, where optical nonlinearities in the cavity for the forward transmission become dominant. By overlapping the bistability curves for the forward and backward directions, one can extract the NTR and power range (PR) of the stable region, where the forward (backward) transmission always follows the upper (lower) branch, as illustrated in Fig. 3(b). Within the stable region, the NTR is rather constant, displaying only a small reduction with input power.

Fig. 3(b) shows that the optimum NTR within the stable region first increases and then decreases with detuning, which is a consequence of the Fano spectral profile, while the PR decreases almost monotonically. When the signal is detuned by a small amount to the transmission minimum of the Fano line (about -3 nm with respect to the resonance wavelength), a large NTR of 35 dB with a PR of 7.8 dB can be achieved. In particular, the device exhibits a NTR larger than 30 dB with a PR varying between 7.5 and 8.5 dB in a wavelength range of more than 1 nm, in which the light power at the input fiber is only ~1 dBm (-6.5 dBm in the input waveguide, as obtained by subtracting the coupling loss of 7.5 dB), cf. Fig. 3(c). Moreover, the operation range increases to 3 nm if an NTR larger than 20 dB suffices. In Fig. 3(b), a PR of more than 10 dB can be achieved together with a NTR between 9.8 and 22.5 dB, and an input power between -10.5 and -2.9 dBm (-17.5 and -10.4 dBm in the input waveguide) in a 1.2 nm wavelength range. The insertion loss is also measured for forward transmission in the stable region, corresponding to the upper branch in the bistability curve (cf. Fig. 3(c)). Since a higher input power is required to switch-on the signal at a larger detuning, the insertion loss increases with detuning, which is consistent with the saturation effect observed in Fig. 2. We find that a larger NTR can be achieved by merely increasing the ratio of $\gamma_1/\gamma_2$, which can be easily achieved by increasing the radius $R_B$ of the blockade hole. In addition, the NTR may be further increased by cascading the Fano structure with other cavity structures, as in Ref. [14], albeit at the prize of increased device complexity.

**Characterization of dynamical properties**

Next, we investigate the nonreciprocal response of the PhC structure when excited by a train of optical pulses, as may be the case for practical applications. Due to self-amplitude modulation, an injected signal will experience a larger nonlinear transmission change at the pulse peak compared to the pulse tail, and may thus efficiently switch on itself even for very small pulse energies. The experimental setup is similar to that in Ref. [31]. A 10 GHz return-to-zero (RZ) signal with 8% duty cycle is on-off modulated at 10 Gbit/s using a pulse pattern generator. Afterwards, the modulated signal is coupled in and out of the PhC device with its polarization aligned to the TE mode. At the output of the device, the signal was first amplified by a standard in-line erbium doped fiber amplifier before being detected with a photodiode with a bandwidth of 45 GHz and monitored with a sampling oscilloscope with a bandwidth of 70 GHz. The signal is spectrally positioned close to the spectral dip of the Fano resonance.

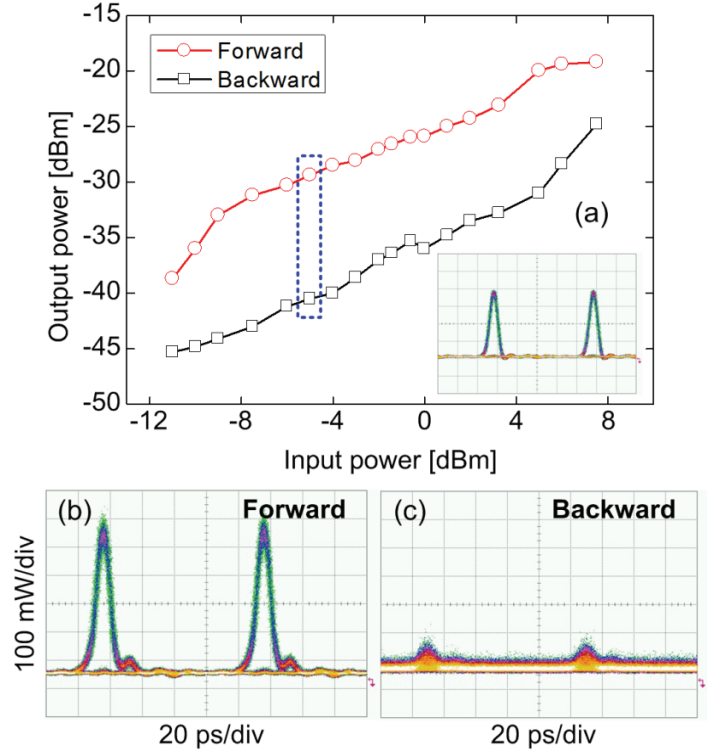

**Figure 4 Nonreciprocal transmission for modulated input signal.** (**a**) Measured averaged output power versus input power (at the input fiber) for forward (red line with circle markers) and backward (black line with square markers) transmission at 10 Gbit/s. The inset shows the eye diagram of the input signal. Eye diagrams of the output signal are shown for (**b**) forward and (**c**) backward transmission when the input power is -5 dBm, corresponding to the dashed blue rectangle in (**a**).

Fig. 4(a) shows the measured averaged output power versus input power for forward and backward transmissions. For forward transmission, the output power first increases relatively quickly with the input power due to the onset of the nonlinearity, and then saturates for higher input powers, ascribed to the increased nonlinear losses. For backward transmission, the output power increases slowly with input power for small input power levels, but the slope of the power curve gradually increases. We find that an NTR of more than 10 dB can be achieved with a PR of 15 dB. Figs. 4(b) and 4(c) show examples of the output eye diagrams at the input power of -5 dBm, exhibiting an extinction ratios of 18.3 and (1.3) for the forward and (backward) transmission, respectively. The input pulse energy in the waveguide is as low as 11 fJ/bit, and the lowest energy for 10 dB NTR is only 4.5 fJ/bit. The clear and open eye observed in the forward direction implies negligible patterning effects, enabled by the fast relaxation of the cavity resonance. The shortest time constant is estimated to be several picoseconds, dominated by carrier diffusion, while a longer time constant of the order of tens to several hundred picoseconds, governed by carrier recombination, determines the slowest response [27-29]. In contrast, the use of thermo-optic nonlinear effects with time constants on the order of a microsecond limits the dynamical bandwidth to the megahertz range. Based on these measurement results, we anticipate that the device can be operated at even larger speed. In Figs. 4(b) and 4(c), a small transient peak is observed following the main peak of the output pulse. This small peak is due to interference between the input signal and the cavity mode [27, 32].

It should be noted that although a large NTR may also be obtained when injecting a high-bit rate signal in a device based on thermo-optic effects [34], this mode of operation is fundamentally different. In that case, the slow thermal relaxation time implies a long-term averaging over the input signal, and the forward signal thus makes the device transparent to any backward propagating signal, ruining the nonreciprocal diode effect. In contrast, the structure we propose is fast enough to block backward transmitting pulses reaching the device in the time interval between two neighboring forward transmitted pulses (see detailed discussions in the supplementary information, section B2). Nevertheless, it remains a challenge to implement a nonreciprocal structure relying on optical nonlinearities that work as an optical isolator for all combinations of forward and backward propagating signals. Besides isolator applications, the suggested structure may act as an intra-cavity saturable absorption element for unidirectional ring sources. We notice that other types of cavities, such as those sustaining whispering gallery modes [33] or interferometer configurations, may be used to eliminate the (constant) back-reflection due to the nanocavity, or alternatively this reflection may be part of the design, e.g. the laser to be isolated. In addition, one may also realize transistor functionality using this structure by applying a relatively weak signal from port 1 having a high coupling rate to the cavity, to switch (or control) a stronger signal coupled simultaneously from port 2 having a lower coupling rate to the cavity.

**Conclusion**

In conclusion, a simple and ultracompact two-dimensional PhC cavity-waveguide membrane structure exhibiting non-reciprocal transmission was proposed and experimentally demonstrated. The structure relies on spatial symmetry breaking in combination with a Fano resonance and the use of cavity-enhanced ultrafast carrier dynamics to realize low-power and high-speed diode characteristics. In comparison to previous works utilizing cascaded cavity structures, our device contains only a single H0 nanocavity, features a very small footprint (< 20 $\mu m^2$), and does not rely on spectral matching of several resonances, which apart from being very difficult limits the operation bandwidth. A Fano transmission spectrum is realized by introducing a single air hole in the waveguide, offering a sharp transmission change within a narrow wavelength range, which enables the simultaneous achievement of a large nonreciprocal transmission ratio (35 dB) and ultra-low threshold power (-6.5 dBm) in a large wavelength range. We further demonstrated the realization of non-reciprocal diode characteristics at high-speed (10 Gbit/s) with ultra-low energy consumption (4.5 fJ/bit). The basic structure demonstrated here allows for further engineering and is expected to be an important building block for realizing integrated isolators or transistors in high-density photonic integrated circuits.

**Associated content**

Further Supporting Information is included: Linear and nonlinear characteristics of the asymmetric cavity-waveguide system.

**Acknowledgment**

We thank M. H. Pu for helpful discussions. The authors acknowledge financial support from Villum Fonden via the NATEC (NAnophotonics for TErabit Communications) Centre.